# Performance Study on Image Encryption Schemes

Jolly Shah and Dr. Vikas Saxena

**Department of CS & IT, Jaypee Institute of Information Technology**
**Noida, Uttar Pradesh 201307, India**

**Department of CS & IT, Jaypee Institute of Information Technology**
**Noida, Uttar Pradesh 201307, India**

**Abstract**
Image applications have been increasing in recent years. Encryption is used to provide the security needed for image applications. In this paper, we classify various image encryption schemes and analyze them with respect to various parameters like tunability, visual degradation, compression friendliness, format compliance, encryption ratio, speed, and cryptographic security.

## 1. Introduction

The use of image and video applications has increased dramatically in recent years. When communication bandwidth or storage is limited, data is often compressed. Especially when wireless network is used, low bit rate compression algorithms are needed because of limited bandwidth. On the other hand, encryption operation is also performed if it is necessary to protect information. Traditionally, an appropriate compression algorithm is applied to multimedia data and its output is encrypted by an independent encryption algorithm. This process must then be reversed by decoder. The processing time for encryption and decryption is a major bottleneck in real time image communication and processing. Along with that processing time required for compression and decompression is also important. The computational overhead incurred by encryption and decryption algorithms makes it impossible to handle tremendous amount of data processed [1], [2]. Encrypting the whole compressed bitstream is very expensive both in delay and processing time, it is proposed in literature to only partially encrypt the compressed bitstream, as a matter of fact, although a large portion of the compressed data is left unencrypted, an adequate choice of bits to encrypt still makes it sufficiently difficult to recover the original data without deciphering the encrypted part so that the security of transmission is achieved [2]. The security of digital images has become more and more important due to the rapid evolution of the Internet in the digital world today. The security of digital images has attracted more attention recently, and many different image encryption methods have been proposed to enhance security of images [4].Image encryption techniques try to convert an image to another one that is hard to understand. On the other hand, image decryption retrieves the original image from the encrypted one. There are various image encryption systems to encrypt and decrypt data.

The rest of the paper is organized as follows: In Section 2 we describe performance parameters based on which image encryption schemes can be compared or evaluated. In section 3, classification and description of these schemes are given. In section 4, performance and comparison of these schemes has been given and finally conclusion is drawn in section 5.

## 2. Performance Parameters

We need to define a set of parameters based on which we can evaluate and compare image encryption schemes. Some parameters listed below are gathered from literature.

**Tunability (T):** It could be very desirable to be able to dynamically define the encrypted part and the encryption parameters with respect to different applications and requirements. Static definition of encrypted part and encrypted parameters limits the usability of the scheme to a restricted set of applications.

**Visual Degradation (VD):** This criterion measures the perceptual distortion of the image data with respect to the plain image. In some applications, it could be desirable to achieve enough visual degradation, so that an attacker would still understand the content but prefer to pay to access the unencrypted content. However, for sensitive data, high visual degradation could be desirable to completely disguise the visual content.

**Compression Friendliness (CF):** An encryption scheme is considered compression friendly if it has no or very little impact on data compression efficiency. Some encryption schemes impact data compressibility or introduce additional data that is necessary for decryption. It is desirable that size of encrypted data should not increase.

**Format Compliance (FC):** The encrypted bit stream should be compliant with the compressor. And standard decoder should be able to decode the encrypted bit stream without decryption.

**Encryption Ratio (ER):** This criterion measures the amount of data to be encrypted. Encryption ratio has to be minimized to reduce computational complexity.

**Speed (S):** In many real-time applications, it is important that the encryption and decryption algorithms are fast enough to meet real time requirements.

**Cryptographic Security (CS):** Cryptographic security defines whether encryption scheme is secure against brute force and different plaintext-ciphertext attack? For highly valuable multimedia application, it is really important that the encryption scheme should satisfy cryptographic security. In our analysis we measure cryptographic security in three levels: low, medium and high.

## 3. Classification and Description of Image Encryption Schemes

In this section, we classify image encryption schemes in two categories. a. Spatial Domain Schemes b. Frequency Domain Schemes. Section 3.1 describes spatial domain image encryption schemes and section 3.2 presents image encryption schemes in frequency domain.

3.1 Spatial Domain Schemes

3.1.1 Methodology proposed by Cheng and Li (2002)

The author proposed a novel solution called partial encryption [5], in which a secure encryption algorithm is used to encrypt only part of compressed data. They proposed partial encryption for quadtree compression. It allows the encryption and decryption time to be significantly reduced without affecting the compression performance of the underlying compression algorithm. In this scheme, the compression output is partitioned into two parts; one is important and other is unimportant parts. Important parts provide a significant amount of information about original data, whereas remaining part called unimportant parts may not provide much information without important parts. Encryption will only perform for important parts. A significant reduction in encryption and decryption time is achieved when the relative size of important part is small.

This scheme is not tunable as static parameters are encrypted. High visual degradation can be achieved only with image having high information rate. As encryption is performed after compression, so no impact is observed on compression efficiency. Encryption ratio can vary from 14% to 50%. Brute force attack is possible for low information images where quadtree structure is very simple. So the security level of this scheme is low.

3.1.2 Methodology proposed by Podesser, Schmitdt and Uhl (2002)

In [6], selective bitplane encryption using AES is proposed. Several experiments were conducted on 8 bit grayscale images, and the main results retained are following: 1. encrypting only the MSB is not secure; a replacement attack is possible 2. Encrypting the first two MSBs gives hard visual degradation, and 3. Encrypting three bitplanes gives very hard visual degradation.

This scheme is not tunable as fix number of bits are encrypted. For 8 bits per pixel uncompressed image, hard visual degradation (of 9 d B) can be observed for a minimum of 3MSB bits encrypted. This scheme is intended for uncompressed data. Encryption can increase data size so it is not compression friendly. In this scheme encryption is performed before compression, so it is format compliant. At least 3 bit planes over 8 (more than 37.5%) of the bit stream have to be encrypted using AES to achieve sufficient security. even when a secure cipher is used (AES), the selective encryption algorithm proposed is vulnerable to replacement attacks [6]. This attack does not break AES but replaces the encrypted data with an intelligible one. It is worth to note that visual distortion is a subjective criterion and does not allow to measure security as illustrated in this example. Security level of this technique can be scaled as medium.

3.1.3 Methodology proposed by Droogenbroeck and Benedett (2002)

This method [4] is proposed for uncompressed image, which applies to a binary image, consist in mixing image data and a message (key) that has the same size as the image: a XOR function is sufficient when the message is only used once. A generalization to gray level images is straightforward: Encrypt each bitplane separately and reconstruct gray level image. With this approach no distinction between bitplanes is introduced although the subjective relevance of each bitplane is not equal. The highest bitplanes exhibit some similarities with the gray level image, but the least significant bitplanes look random. Because encrypted bits also look random, the encryption of least significant bitplanes will add noise to the image. The advantage of least significant bits is that plaintext attacks are harder on random like data. It is preferable to encrypt bits that look most random.

This scheme is tunable. Very high visual degradation can be achieved by encrypting 4 to 5 bitplanes. This technique is used for uncompress image so no impact is observed on compression efficiency. In this scheme encryption ratio vary from 50 to 60%. It is fast as XOR operation takes less time. It is not robust against cryptanalysis attack. So, security level is low.

3.1.4 Methodology proposed by Y.V.Subba Rao, Abhijit Mitra and S.R.Mahadeva Prasanna (2006)

In the proposed scheme [7], the image is initially separated into correlated and uncorrelated data by dividing it into first four MSB planes and last four LSB planes. The correlated data are encrypted with highly uncorrelated pseudo random sequence while keeping the uncorrelated data as unencrypted ones. After encryption of correlated data it combines with uncorrelated data to form final encrypted image. After the encryption of the first two MSB planes the encrypted image appears as noisy image. However, it is also necessary to encrypt third and fourth bitplane because these bitplanes have some perceptual information. The resultant image does not have significant information after the encryption of four MSB planes.

This scheme is tunable. By encrypting first four MSB plane high visual degradation can be achieved. No impact is observed on compression efficiency. In this scheme 50% of bits are encrypted. Speed is fast.

3.1.5 Methodology proposed by Nidhi s. Kulkarni, Indra Gupta and Shailendra N. Kulkarni (2008)

The proposed encryption technique [8] reduces intelligent information in an image by scrambling the image first and then changing the pixel values. The scrambling arrangement is done with the help of a random vector and the pixel values are changed by a simple substitution method which adds confusion and diffusion property to encryption technique. The proposed technique has advantage of convenient realization, less computational complexity and better security. The algorithm is suitable for any kind and any size of gray level image.

Tunability of this scheme is not specified. High visual degradation can be achieved. It is compression friendly as well as speed is fast. Encryption ratio of this scheme is 100% as all bits are encrypted. Moderate level of security is obtained. There is a scope of improvement to increase security against brute force attack.

3.1.6 Methodology proposed by Alireza Jolfaei and Abdolrasoul Mirghadri (2010)

In this scheme [9], pixel shuffler and stream cipher unit are used to encrypt an image. Pixel scrambling has two important issues that are useful for image ciphering. It not only rearranges pixel location but also changes the value of each pixel. Confusion is performed by stream cipher itself through nonlinear function operation. Pixel location displacement is appropriate before applying encryption, because unlike the text data that has only two neighbours, each pixel in the image is in neighbourhood with eight adjacent pixels. For this reason, each pixel has a lot of correlation with its adjacent neighbours. However, it is very important to disturb the high correlation among image pixels to increase the security level of encrypted images. In order to dissipate the high correlation among pixels, pixel shuffler is used; in which permutation map is applied in two directions: vertical and horizontal, to decrease adjacent pixel correlation. The proposed scheme's key space is large enough to resist all kinds of brute force attack.

This scheme is not tunable. High visual degradation can be achieved. No impact is observed on compression efficiency. Encryption ratio is 100%. Speed is fast. It is robust against brute force attack so security level is high.

3.2 Frequency Domain Schemes

3.2.1 Methodology proposed by Cheng and Li (2002)

The wavelet-based compression algorithm SPIHT partitions the data into two parts. The first part can be considered as the "important part", it consists of significant information of coefficients and sets for the two highest levels of the pyramid and the initial threshold parameter n of significance computation ($T^n$). The second part is the "unimportant part"; it consists of sign bits and refinement bits. No encryption algorithm is specified, only the important part is encrypted [5].

This scheme is not tunable as static parameters are encypted. High visual degradation can be achieved. No impact is observed on compression efficiency. SPIHT is not part of any compression standard. In additions, since SPIHT algorithm is context based, no decoding/processing is possible without the knowledge of the first significant bits. It is not format compliant. Due to energy concentration obtained by DWT only 7% of the bitstream is encrypted. As encryption algorithm is not specified, so it is difficult to determine speed of this scheme. Speed of this scheme varies based on encryption algorithm that it uses to encrypt important part. If the two highest resolutions are very small, brute force attack becomes possible to guess the initial threshold and significance information. Security is Moderate.

3.2.2 Methodology proposed by Droogenbroeck and Benedett, (2002)

This method [4] is a selective encryption for JPEG image which is based on the encryption of DCT coefficients. In this method JPEG Huffman coder terminates runs of zeros with codewords /symbols in order to approach the entropy. Appended bits are added to these codewords to fully specify the magnitudes and signs of nonzero coefficients, only these appended bits are encrypted using DES or IDEA.

This scheme is not tunable. High visual degradation is achievable. The encryption is separated from Huffman coder and has no impact on the compression efficiency. It is JPEG compliant. Very high encryption ration is required about 92%. About 92% of the data is encrypted using well scrutinized symmetric ciphers. It would be very difficult to break the encryption algorithm or try to predict the encrypted part. High security can be obtained by use of standard symmetric cipher.

3.2.3 Methodology proposed by Zeng and Lei (2003)

In [10], selective encryption in the frequency domain ($8 \times 8$ DCT and wavelet domains) is proposed. The general scheme consists of selective scrambling of coefficients by using different primitives such as selective bit scrambling, block shuffling, and/or rotation. In wavelet transform case selective bit scrambling and block shuffling is done. In selective bit scrambling the first nonzero magnitude bit and all subsequent zero bits if any give a range for the coefficient value. These bits have low entropy and thus highly compressible and all remaining bits called refinement bits are uncorrelated with the neighbouring coefficients. In this scheme, signbits and refinement bits are scrambled. In block shuffling, the basic idea is to shuffle the arrangement of coefficients within a block in a way to preserve some spatial correlation; this can achieve sufficient security without compromising compression efficiency. Each subband is split into equal-sized blocks. Within the same subband, block coefficients are shuffled according to a shuffling table generated using a secret key. Since the shuffling is block based, it is expected that most 2D local subband statistics are preserved and compression not greatly impacted.

In DCT transform case, the $8 \times 8$ DCT coefficients can be considered as individual local frequency components located at some subband. The block shuffling and sign bits change can be applied on these "subbands." I, B, and P frames are processed in different manners. For I-frames, the image is first split into segments of macroblocks, blocks/macroblocks of a segment can be spatially disjoint and chosen at random spatial positions within the frame. Within each segment, DCT coefficients at the same frequency location are shuffled together. Then, sign bits of AC coefficients and DC coefficients are randomly changed. There may be many intracoded blocks in P- and B-frames. At least DCT coefficients of the same intracoded block in P- or B-frames are shuffled. Sign bits of motion vectors are also scrambled.

It is not tunable. High-visual degradation is achieved. Indeed, most of the image energy is concentrated in DC coefficients, thus, encrypting them affects considerably the image content. It is not compression friendly as bitrate increase by about 20% is observed. It is compliant with JPEG and MPEG standards. If we consider only the AC sign bit encryption, it represents 16 to 20% of data. This is relatively high. It is difficult to define speed of this scheme; as encryption algorithm is not specified. So speed of this scheme generally varies based on encryption algorithm in use. It is vulnerable to chosen and known plaintext attacks since it is based only on permutations. In addition, replacing the DC coefficients with a fixed value still gives an intelligible version of the image.

3.2.4 Methodology proposed by Pommer and Uhl (2003)

The authors proposed wavelet packet based compression instead of pyramidal compression schemes in order to provide confidentiality [11]. Header information of a wavelet packet image coding scheme that is based on either a uniform scalar quantizer or zerotrees is protected: it uses AES to encrypt only the subband decomposition structure. In this approach the encoder uses different decomposition schemes with respect to the wavelet packet subband structure for each image. it is based on AES encryption of the header information of wavelet packet encoding of an image, this header specifies the sub band tree structure. These decomposition trees are encrypted and have to be present at the decoder to be able to reconstruct the image data properly. The advantage in comparison to other selective encryption approaches is that the amount of necessary encryption is extremely small since only header information, and no visual data, needs to be processed.

It is not tunable. The encrypted content cannot be viewed without decryption. The subband tree is pseudo randomly generated. This adversely impacts the compression efficiency. It is not format compliant. The encrypted part represents a very small fraction of the bitstream. It is not secure against chosen plaintext attack. Because statistical properties of wavelet coefficients are preserved by the encryption, then the approximation subband can be reconstructed. This will give the attacker the size of the approximation subband (lower resolution) and then neighbouring sub bands can be reconstructed since close subbands contain highly correlated coefficients.

3.2.5 Methodology proposed by Han Shuihua and Yang Shuangyuan (2005)

In [13], a novel asymmetric image encryption scheme is proposed. In this scheme, based on certain matrix transformation, all the pixels and frequencies in each block of original image are scrambled. To implement this algorithm, first, a pair of keys is created based on matrix transformation; second, the image is encrypted by using private key in its transformation domain; finally the receiver uses the public key to decrypt the encrypted image. Because it is based on matrix transformation, it is easily implemented and highly efficient to quickly encrypt and decrypt image messages.

It is not tunable. High visual degradation can be achieved. It affects compression efficiency. It is JPEG compliant. Encryption ratio is 100%. Speed is moderate. It is not secure against all cryptanalysis attack; security has room to improve. So, security level is moderate.

3.2.6 Methodology proposed by Omar M. Odibat, Moussa H. Abdallah and Moh'd Belal R. Al-Zoubi (2006)

In this paper [14], multilevel partial image encryption method has been proposed in which image encryption is performed before compression. Encryption is performed on low frequency coefficients. Then discrete Fourier transform is applied on the approximation coefficients only. The result will be permuted using a permutations matrix, which serves as the encryption key. The permutation step is followed by wavelet reconstruction; this will generate the encrypted image. In final stage, compression is done using Huffman coding. This scheme is flexible as various digital image transformations could be used instead of DFT. In addition, any scaling function could be applied. And finally any image compression techniques could be used.

It is not tunable. Low visual degradation can be achieved. It is compression friendly. Encryption Ratio is 1.56%. Speed is moderate. It is not secure against replacement attack. If an attacker knows the unencrypted parts, which are high frequency coefficients in frequency domain, some outlines of original image may appear.

3.2.7 Lala Krikor, Sami Baba, Thawar Arif, and Zyad Shaaban (2009)

The proposed method [12] is based on the idea of decomposing the image into 8X8 blocks; these blocks are transformed from the spatial domain to frequency domain by the DCT. Then, the DCT coefficients correlated to the higher frequencies of the image block are encrypted using Non-linear Shift Back Register. The concept behind encrypting only some selective DCT coefficients is based on the fact that the image details are situated in the higher frequencies while the human eye is most sensitive to lower frequencies than to higher frequency. The proposed algorithm is lossless; hence the images used in such applications are of highly important information, and any amount of information is not allowed.

It is tunable as different level of security can be achieved by selecting different bits for encryption. In this scheme, variable visual degradation can be achieved. No impact is observed on compression efficiency so it is compression friendly. Encryption ratio is variable. To increase security, block shuffling methods is applied after encryption.

3.2.8 Methodology proposed by Shaimaa A. El-said, Khalid F. A. Hussein and Mohamed M. Fouad (2010)

In [15], a secure and computationally feasible algorithm called optimized multiple Huffman table is proposed. OMHT depends on using statistical model based compression method to generate different tables from the same data type of images or videos to be encrypted leading to increase compression efficiency and security of the used tables.

High visual degradation can be achieved. No impact is observed on compression efficiency. Encryption ratio is 100%. It is resistant against various types of attack including cipher text only attack and known/chosen plaintext attack.

## 4. Comparison of Image Encryption Schemes

This section presents performance and comparison among image encryption schemes with respect to various parameters as shown in Table 1.

The main symbol use is: "?" for unspecified criteria. We specify tunability as either "yes' or "no". Visual degradation is measured in three levels: high, low, variable. It is desirable that visual degradation is variable and dynamically tunable to adapt to different application

Table1. Performance Comparison of Image Encryption Schemes

| Type of Data | Domain | Proposal | Encryption Algorithm | Tunable | Visual Degaradation | Compression Friendliness | Format Compliance | Encryption Ratio | Speed | Cryptographic Security | What is Encrypted? |
|---|---|---|---|---|---|---|---|---|---|---|---|
| Image | Spatial Domain Schemes | Cheng and Li (2002)[5] | Not specified | no | High | yes | not applicable | 14%-50% | Fast | low | Quadtree structure |
| | | Podesser, Schmidt and Uhl (2002)[6] | AES | no | High | no | yes | >37.5% | Fast | medium | First three MSB planes |
| | | Droogenbroeck and Benedett (2002)[4] | XOR | yes | High | yes | not applicable | 50-60% | Fast | low | Least Significant 4-5 bitplanes |
| | | Y.V.Subba Rao, Abhijit Mitra and S.R.Mahadeva Prasanna(2006)[7] | PsuedoRandom Sequence | yes | High | no | ? | ≈50% | Fast | ? | First four MSB planes |
| | | Nidhi s. Kulkarni, Indra Gupta and Shailendra N. Kulkarni (2008)[8] | Shuffling using Random Vector | ? | High | yes | ? | 100% | Fast | Moderate | Complete Image Matrix |
| | | Alireza Jolfaei and Abdolrasoul Mirghadri (2010)[9] | Pixel Shuffler/Stream Cipher (XOR with Keystream) | no | High | yes | ? | 100% | Fast | High | Complete Image Matrix |
| | Frequency Domain Schemes | Cheng and Li(2002)[5] | no algorithm specified | no | High | yes | no | 7% | Variable | Moderate | Pixel and set related significance information in the two highest pyramid level of SPIHT |
| | | Droogenbroeck and Benedett (2002)[4] | DES,IDEA | no | High | yes | JPEG | 92% | slow | High | sign and magnitude of non zero DC |
| | | Zeng and Lei(2003)[10] | no algorithm specified | no | High | yes | complaint DWT/ JPEG/MPEG (DCT domain) | 20% | variable | Low | sign bits/refinement bits(DWT), DC,AC,Motion vectors(DCT) |
| | | Pommer and Uhl (2003)[11] | AES | no | High | no | no | small | ? | Low | Header |
| | | Han Shuihua and Yang Shuangyuan (2005)[13] | Matrix transformation | no | High | no | JPEG | 100% | moderate | moderate | Each block of image |
| | | Omar M. Odibat, Moussa H. Abdallah and Moh'd Belal R. Al-Zoubi (2006)[14] | not specified | no | low | yes | ? | 1.56% | moderate | low | low frequency coefficients |
| | | Lala Krikor, Sami Baba, Thawar Arif, and Zyad Shaaban (2009)[12] | NLFSR(xor)KEY | yes | variable based on selection of DCT coefficients | yes | ? | variable | fast | High if block shuffling is used | Selective DC and AC coefficients |
| | | Shaimaa A. El-said, Khalid F. A. Hussein and Mohamed M. Fouad (2010)[15] | Optimized Multiple Huffman Table | ? | High | yes | ? | 100% | ? | High | Complete Image |

requirements. We specify compression friendliness and format compliance as "yes" or "no". Encryption ratio is specified in "%". It is % of the data that is encrypted with respect to total. Speed is measured as fast, low, moderate, variable. Speed is decided based on underlying encryption algorithm. Cryptographic security is measured in three levels: high, moderate and low. If encryption scheme is robust against all cryptanalysis and special attacks then security of this scheme is measured as high. If it is secure against some of the cryptanalysis attacks then it is measured as moderate. And if it is not secure against any of the cryptanalysis attack then its security is measured as low.

## 5. Conclusion

We have compared spatial domain and frequency domain image encryption schemes. We analyze these schemes with respect to various parameters as shown in table 1. We can conclude that spatial domain schemes are generally faster but security level is low. In spatial domain, the scheme mentioned in [9] is good as far as compression friendliness, security and speed is concerned. While in frequency domain, scheme [4] [12] [15] maintains good performance from security point of view. Security level and speed performance is generally span from low to high for different frequency domain schemes. The encryption scheme proposed in [12] is good as far as security, speed and compression efficiency altogether is concerned. We can conclude that none of the schemes mentioned in table 1 satisfies all performance parameters. So, it is a challenge for a research to design an encryption scheme which maintains good tradeoff among tunability, visual degradation, compression friendliness, format compliance, encryption ratio, speed, and cryptographic security.

**Jolly shah** she completed her undergraduate degree in information Technology at Dharamsinh Desai University in 2004. She obtained her master degree in Computer Engineering from Bharati Vidyapeeth University in 2008. She is pursuing PhD in Computer Science from Jaypee Institute of Information Technology. Her research interest are in cryptography, video processing.

**Dr. Vikas Saxena** he completed his undergraduate degree in Computer Science at at Rohilkhand Univ. He obtained his master degree from VJIT, Mumbai. He completed his PhD in Computer Science from Jaypee Institute of Information Technology. He is recently serving as an assistant professor in JIIT. He published various papers in International journals and conferences. His area of interest is image processing, graphics and software engineering.